\documentclass{aa}
\usepackage{epsfig}

\begin{document}

\title{SPI/INTEGRAL in-flight performance}

\author{
     J. P. Roques \inst{1}, S. Schanne \inst{3}, A. von Kienlin \inst{2}, J. Kn\"odlseder \inst{1},
 R. Briet \inst{10}, L. Bouchet \inst{1}, Ph. Paul \inst{1}, S. Boggs \inst{8}, P. Caraveo \inst{4},
 M. Cass\'e \inst{3}, B. Cordier \inst{3}, R. Diehl \inst{2}, P. Durouchoux \inst{3}, P. Jean \inst{1},
 P. Leleux \inst{6}, G. Lichti \inst{2}, P.~Mandrou \inst{1}, J. Matteson \inst{7},
 F. Sanchez \inst{5}, V. Sch\"onfelder \inst{2}, G. Skinner \inst{1}, A. Strong \inst{2},
 B. Teegarden \inst{9}, G.~Vedrenne \inst{1}, P. von Ballmoos \inst{1},
     \and 
 C. Wunderer \inst{2} 
     }

\authorrunning{Roques et al.}
\offprints{J.P.Roques ; {\em roques@cesr.fr} }

\institute{
    Centre d'Etude Spatiale des Rayonnements, CNRS/UPS, B.P. 4346, 31028 Toulouse, France
\and  
    Max-Planck-Institut f\"ur extraterrestrische Physik, Postfach 1603, 85740 Garching, Germany 
\and
    DSM/DAPNIA/SAp, CEA-Saclay, 91191 Gif-sur-Yvette, France
\and
    IASF, via Bassini 15, 20133 Milano, Italy 
\and
    IFIC, University of Valencia, 50 avenida Dr. Moliner, 46100 Burjassot, Spain
\and
    Institut de Physique Nucl\'eaire, Universit\'e catholique de Louvain, B-1348 Louvain-La Neuve, Belgium
\and
   UCSD/CASS, 9500 Gilman Drive, La Jolla, CA 92093-0111, USA
\and
    Space Science Laboratory, University of California, Berkeley, CA 94720, USA
\and 
    Laboratory for High Energy Astrophysics, NASA/Goddard Space Flight Center, Greenbelt, MD 20771, USA
\and
    CNES/CST,18 avenue Edouard Belin, 31401 Toulouse cedex 4, France
}

\date{Received July 30, 2003; accepted  September 23, 2003}

 \abstract{
The SPI instrument has been launched on-board the INTEGRAL observatory on October 17, 2002. SPI is a spectrometer devoted to the sky observation in the 20 keV-8 MeV energy range using 19 germanium detectors. The performance of the cryogenic system is nominal and allows to cool the 19 kg of germanium down to 85 K with a comfortable margin. The energy resolution of the whole camera is 2.5 keV at 1.1 MeV. This resolution degrades with time due to particle irradiation in space. We show that the annealing process allows the recovery of the initial performance. The anticoincidence shield works as expected, with a low threshold at 75 keV, reducing the GeD background by a factor of 20. The digital front-end electronics system allows the perfect alignement in time of all the signals as well as the optimisation of the dead time (12\%). We demonstrate that SPI is able to map regions as complex as the galactic plane. The obtained spectrum of the Crab nebula validates the present version of our response matrix. The 3 $\sigma$ sensitivity of the instrument at 1 MeV is 8 10$^{-7}$ph$\cdot$cm$^{-2}\cdot$s$^{-1}\cdot$keV$^{-1}$ for the continuum and 3 10$^{-5}$ph$\cdot$cm$^{-2}\cdot$s$^{-1}$ for narrow lines.

\keywords{Gamma-ray : instrument, observations --- Space telescope : INTEGRAL/SPI}
}
\maketitle
\section{Introduction}
The spectrometer SPI consists of the following main subsystems:
\begin{itemize}
\item The camera, composed of 19 high purity germanium detectors (GeDs) and their associated electronics.
\item A two-stage cooling system: the passive stage cools the cryostat housing and the preamplifiers to 215 K; the active stage cools the Ge array down to 85-90 K.
\item A pulse shape discrimination system (PSD) which allows discrimination between single and multi-site interactions in one Ge detector.
\item An active anticoincidence shield (ACS) made of 91 BGO blocks and a plastic scintillator (PSAC).
\item A digital front-end electronics (DFEE) providing: the SPI internal timing, the various (anti)coincidence functions  and the primary data encoding.
\item A coded mask which allows imaging of the sky.
\end{itemize}
A complete description of SPI can be found in \cite{ved2003}.
In this paper we review:
\begin{itemize}
\item The in-flight performance of the SPI subsystems.
\item The tuning of some of the instrument parameters
\item The global SPI performance after tuning
\item The evolution of some instrument characteristics during flight.
\end{itemize}
The results presented here cover 9 months of operation in space (October 2002 - July 2003). This period can be divided in the following phases:
\begin {itemize}
\item The performance verification phase (PV-phase) from 2002 Oct. 17 to Dec. 30 which consists of the first activation, the tuning, the performance measurements and the scientific validation of the instrument.
\item The first science observation phase from 2002 Dec. 30 to 2003 Feb. 5.
\item The first annealing from 2003 Feb. 5 to 18
\item The second science observation phase from 2003 Feb. 18 to 2003 July 18.
\end {itemize}
\section{Camera thermal performance}
      \subsection{Cryostat outgassing and cooling phase}
\begin{figure} [here]
\vspace*{-0.5cm}
 \centering
  \includegraphics[width=6.0cm,angle=270]{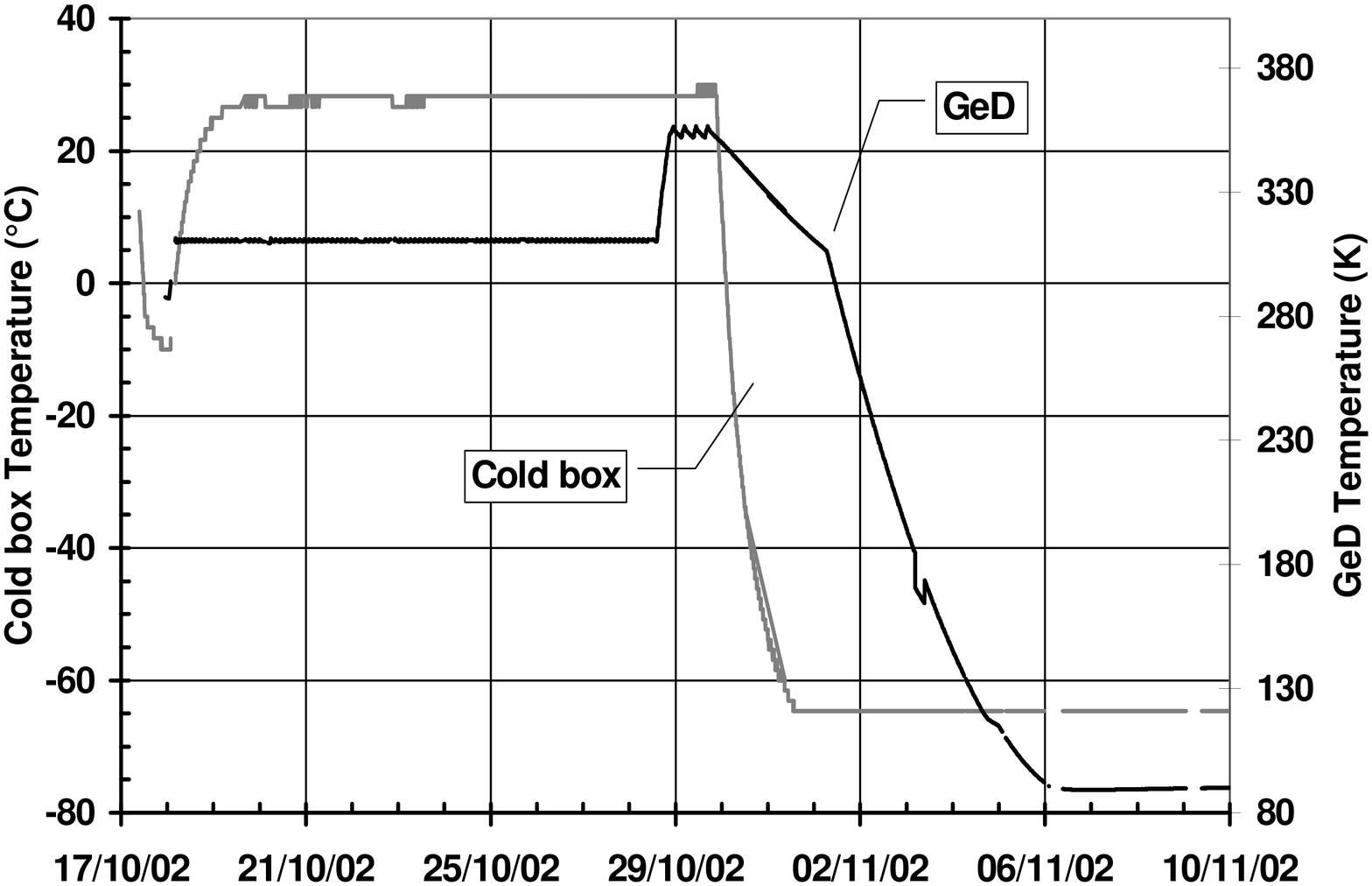}
\caption{GeD and cold box temperature profile.}
\label {therm1fig}
\end{figure}

The initial activation of the camera has been performed in several steps, summarised in figure \ref {therm1fig}. This figure shows the cold plate (i.e. the Ge detectors) temperature and the intermediate stage (i.e. Ge detector preamplifiers) temperature versus time. The following steps were performed:

\begin{itemize}
\item

Cryostat outgassing: 16 hours after launch the outgassing started, the GeDs were maintained at 37.8$^\circ$C and the cold box at 28$^\circ$C during 10 days. The temperature of the GeDs was then increased to 81$^\circ$C for 24 hours. This outgassing scheme is necessary to clean the rear surface of the GeDs. Pollution on this surface dramatically increases the leakage current of the GeDs and reduces their energy resolution. It is important to note that the GeDs are not sealed and that the cryostat was filled with air the last few days before launch.

\item
Cooling phase: after a period of passive cooling the cryocoolers have been switched on. The cold plate reached 90 K on Nov 5th. The intermediate stage reached ${-66}^\circ$C on Nov. 1st.
\end {itemize}

 \subsection {Nominal operations and stability of the cold plate: November 11, 2002-February 5, 2003}

The objective was to maintain the cold plate at 90$\pm$1~K; this was achieved initially by setting the stroke of the stirling machines to 56\% of the maximum. However in the long term the temperature of the cold plate slightly increased and the compressors strokes were increased to compensate this effect. The drifts were:
        Nov. 11 to Nov. 22  +0.1 K/day; 
        Nov. 22 to Dec. 9   +0.06 K/day;
        Dec. 9 to Feb. 6 2003 +0.03 K/day.

This long-term positive temperature drift was due to contamination of the insulation of the cold parts by outgassing of the whole spacecraft. This contamination decreases the efficiency of the insulation. The decrease of the temperature increase rate is consistent with the decrease of the outgassing with time.

\subsection {Nominal operations after annealing}

        The first in-flight annealing process took place from Feb. 6 to 18, 2003, (see annealing section). It was decided to operate the cold plate at 85$\pm$1 K after this process. The cryocoolers settings needed to maintain this temperature were the same as those needed before annealing to maintain 90 K. The thermal load decrease on the coolers has been estimated to be 250 mW. The outgassing process during the annealing has cleaned the cold surfaces. The hypothesis of contamination by the spacecraft outgassing was thus confirmed. Nevertheless this contamination process did not stop, the temperature drift was:
        Feb. 25 to March 7 +0.08 K/day;
        March 8 to May +0.03 K/day and tends to stabilize afterwards.

\subsection {Nominal operations and stability of the intermediate stage}
        This intermediate temperature stage, cooled by amonnia heat pipes and a passive radiator system, relies on thermal regulation in order to operate above the ammonia freezing point. The cryostat envelope is thus maintained at -66$^\circ$C with a non-measurable drift (sensor resolution is 0.5$^\circ$C) within the 9 months of operations.

Each GeD preamplifier, situated inside the cryostat, has a thermal sensor. The temperature stability is excellent and no change can be measured within the temperature resolution which is 0.1$^\circ$C. Thus the gain drifts linked to preamplifier temperature drifts are negligible.

\section{Camera performance measurements}
\subsection{Energy resolution after first camera switch on} 
\begin{figure} [here]
\vspace*{-1cm}
 \centering
  \includegraphics[width=6.5cm,angle=270]{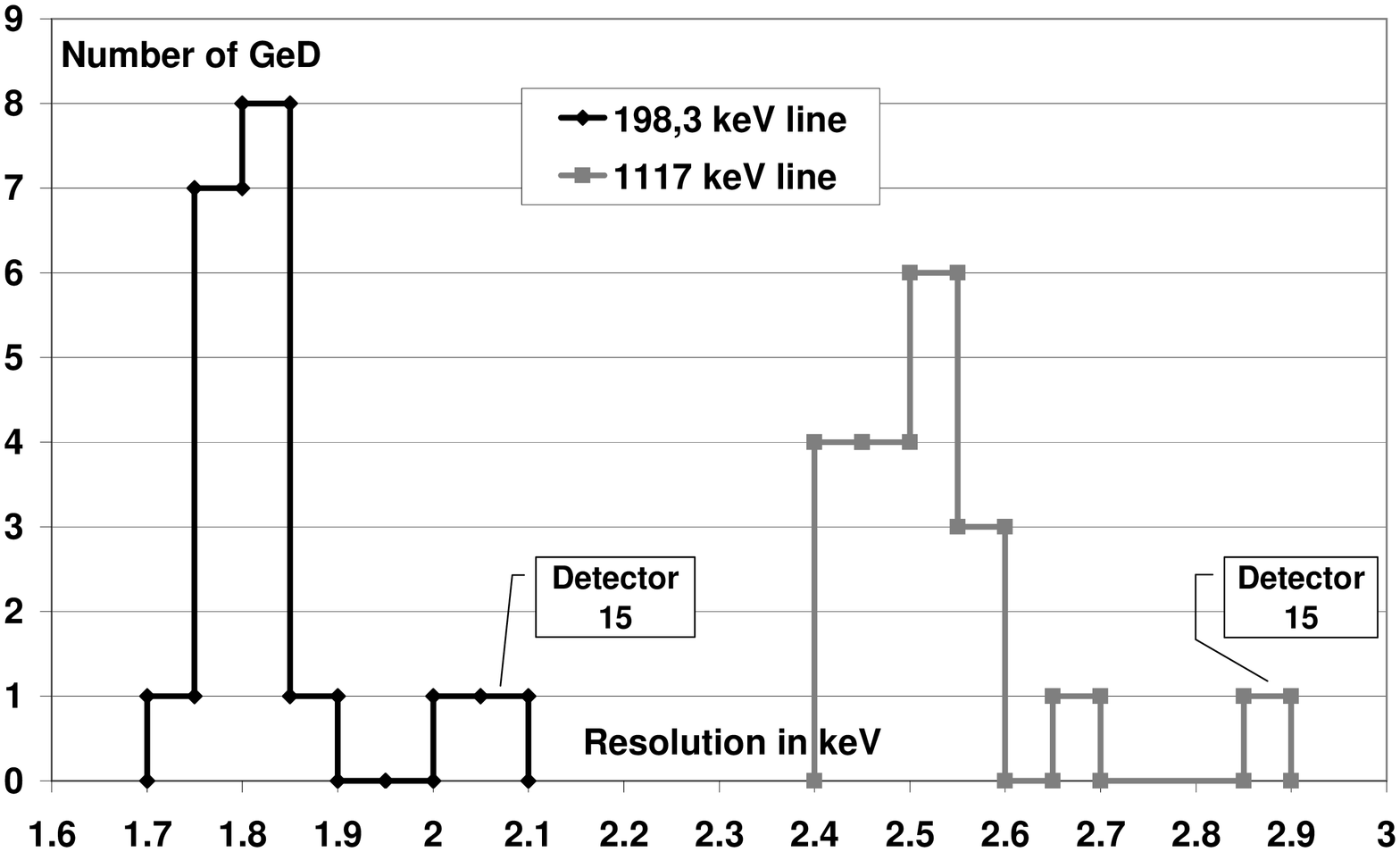}
\vspace*{-0.5cm}
\caption{Number of GeD per energy resolution interval for the 198 and 1117 keV lines.}
\label {distri1fig}
\end{figure}

Shortly after the launch all camera electronics units were switched on with the exception of the GeD high voltages (HV). The HV were applied in steps of 500 V when the GeD temperature reached 117 K on November 4th. We started the GeDs performance measurements at a relatively elevated temperature. In this case, if the GeDs are polluted, the resulting leakage current increase (which affects the energy resolution) is much higher.
Already the first set of measurements indicated clearly that GeD 15 exhibited an abnormal behaviour, indicating an important leakage current, increasing with the high voltage, probably due to contamination. Continuous measurements were performed down to the 90 K target temperature.
Figure \ref {distri1fig} gives the energy resolution at 90 K of all the GeDs for 2 background lines: 198.3 keV and 1117.3 keV. If we exclude GeD 15, the results obtained show a good homogeneity of the whole detection plane with a mean energy resolution of 2.5 keV at 1.1 MeV and 1.8 keV  at 198 keV. The GeD HV was set to 4000 V except for GeD 15. For comparison the energy resolution measured on ground was 2.45 keV at 1332 keV.

The energy resolution of GeD 15 improved at 90 K as expected for a contamination problem, nevertheless it was not possible to leave the detector at its nominal HV setting; the best compromise found was to set the HV to 2500 V. With this setting the energy resolution obtained was 2.1 keV at 198 keV and 2.85 keV at 1.1 MeV. This setting was kept until the first annealing.

\subsection {Energy resolution after the first annealing}
The annealing process leads to an important outgassing of the cryostat. After this process, the switch-on procedure performed at 117 K showed a perfect behaviour of all GeDs as a function of the HV, in particular the GeD 15 performance was found to be nominal and the HV of all GeDs have been set to 4000 V since this date. This confirmed that the GeDs were contaminated before this annealing. The cold plate was then cooled down to 85 K. The mean energy resolution of the camera at this temperature was 1.82 keV at 198 keV and 2.9 keV at 1764 keV.
These results show that since this second outgassing (associated with the annealing) all the GeDs are in perfect health.

\subsection {Energy resolution versus energy - Time stability}

\begin{figure}
 \centering
  \includegraphics[width=6.5cm,angle=270]{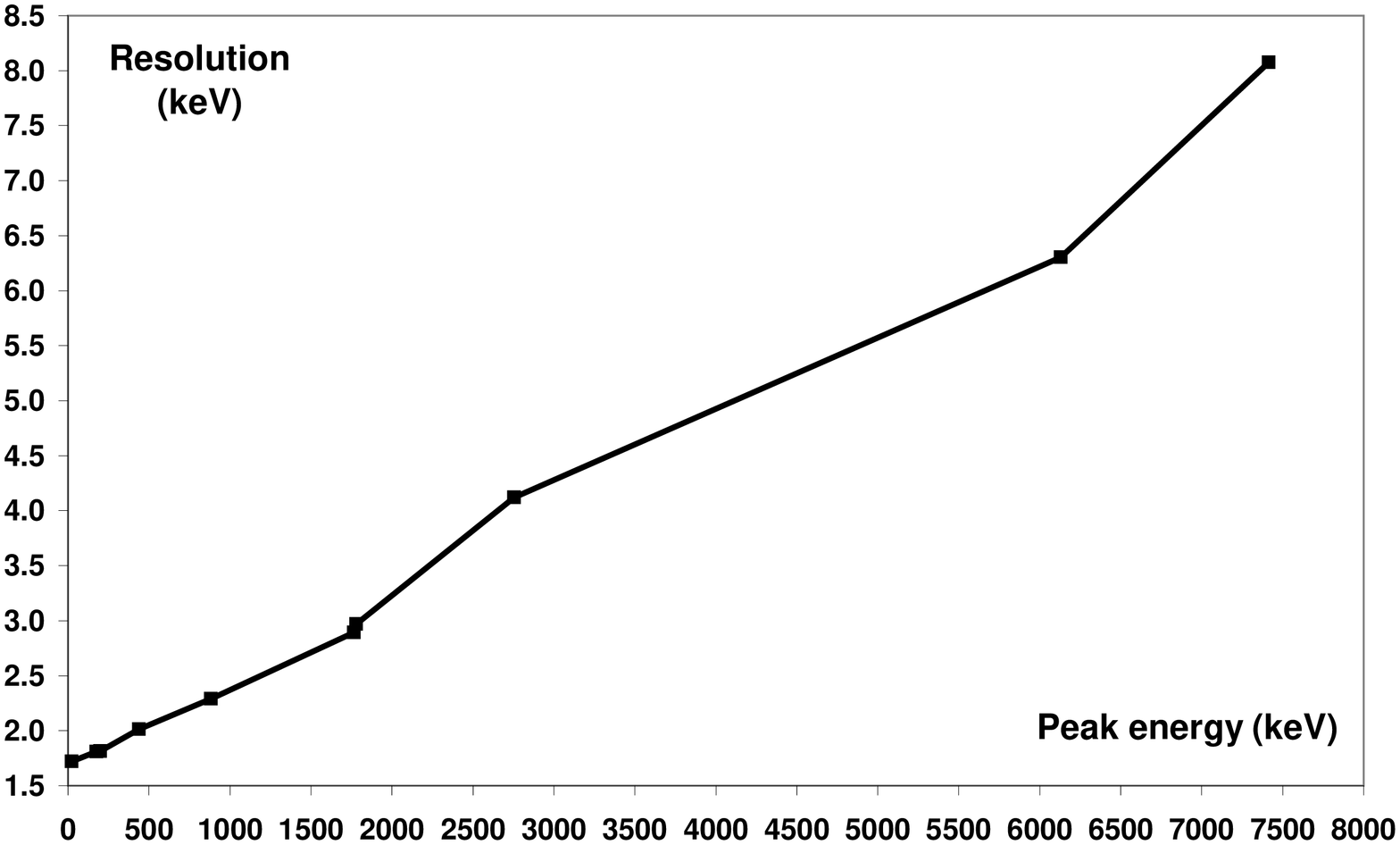}
\vspace*{-0.5cm}
\caption{Mean camera (GeD 15 is excluded) energy resolution in the whole energy range one month after launch.}
\label {allresol}
\end{figure}

The mean energy resolution of the camera over the whole energy range has been determined by simply integrating counts in the 19 energy spectra during 3.8 days one month after launch with stable GeD temperature without applying any gain correction. The energy resolutions have been determined from these spectra for each detector individually and then averaged over all 19 detectors. These results are shown in figure \ref {allresol}. The energy resolution ranges from 1.73 keV to 8 keV. This result demonstrates the perfect stability of the electronics over a long time period.

\subsection {Energy resolution versus irradiation}

\begin{figure}
\centering
  \includegraphics[width=6.0cm,angle=270]{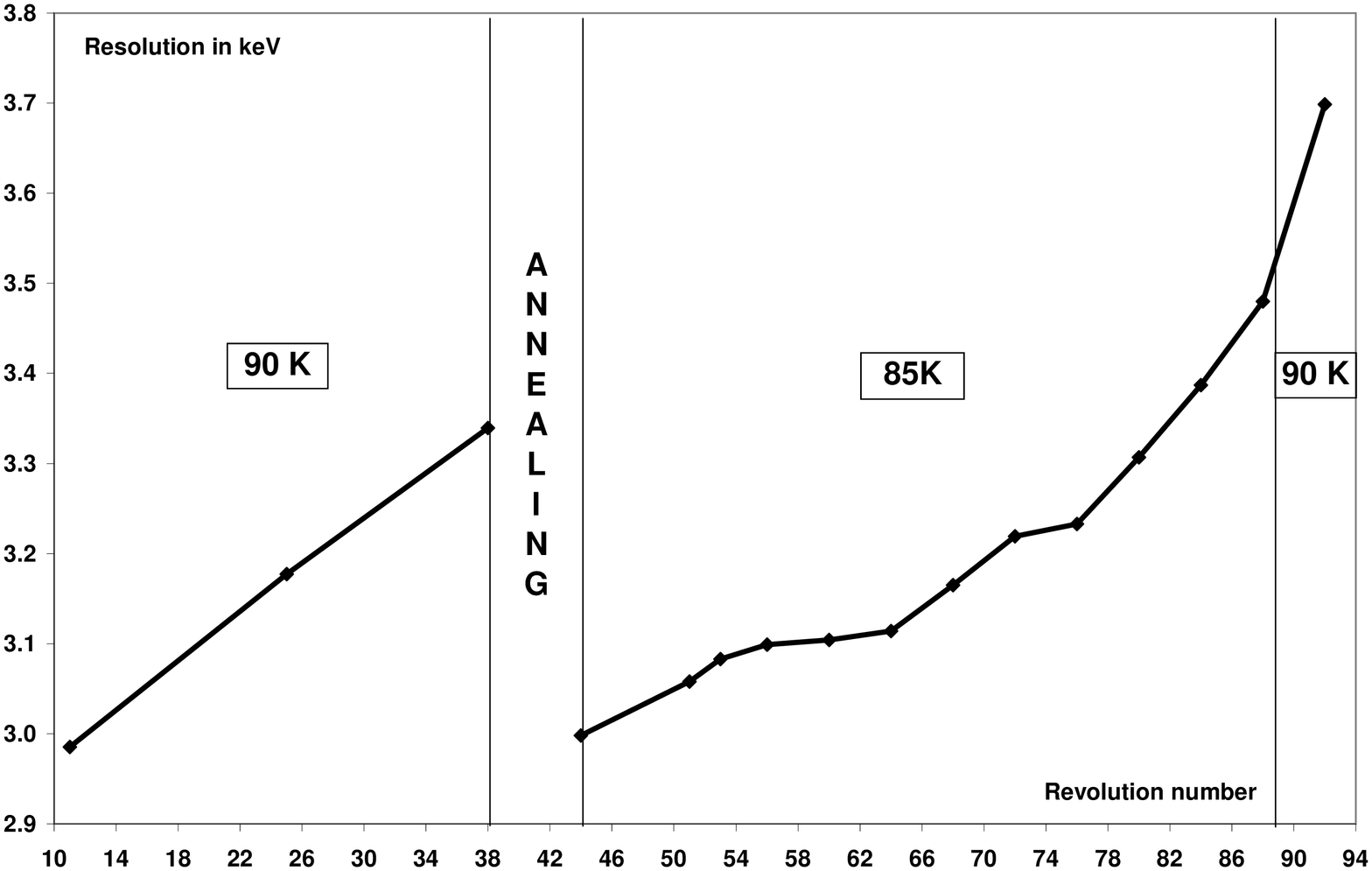}
\vspace*{-0.5cm}
\caption{Energy resolution evolution at 1778 keV.}
\label {resol-ann}
\end{figure}

The GeD performance degrades with the damages created within the crystal by incident radiation (mainly protons and neutrons). Radiation damage increases the number of hole traps within the active detection volume (\cite{knoll}). In space the particule flux  is high enough to produce substantial degradation of the GeDs over times of a few months. Thus energy resolution of the GeDs evolves with time and a continuous monitoring of the resolution is needed to analyse properly the data from SPI.
Figure \ref {resol-ann} represents the mean energy resolution of the camera as a function of time. The first period shows the evolution at a temperature of 90 K. The second period shows energy degradation only ${\sim}$ 0.7 times as fast, thanks to a lower GeD temperature (85 K). Discussion on GeD irradiation and temperature dependance can be found in \cite{mahoney}, \cite{borrel}, \cite{kandel}.

\subsection { Annealing}

The annealing process consists of a baking of the GeDs (100$^\circ$C) in order to restore the quality of the crystalline lattice and suppress the trapping sites. The success of such a process highly depends on the initial damage and on the annealing temperature and duration. SPI GeDs are heated by four resistors glued on the GeD array support.

\begin{figure} [here]
 \centering
  \includegraphics[width=6.0cm,angle=270]{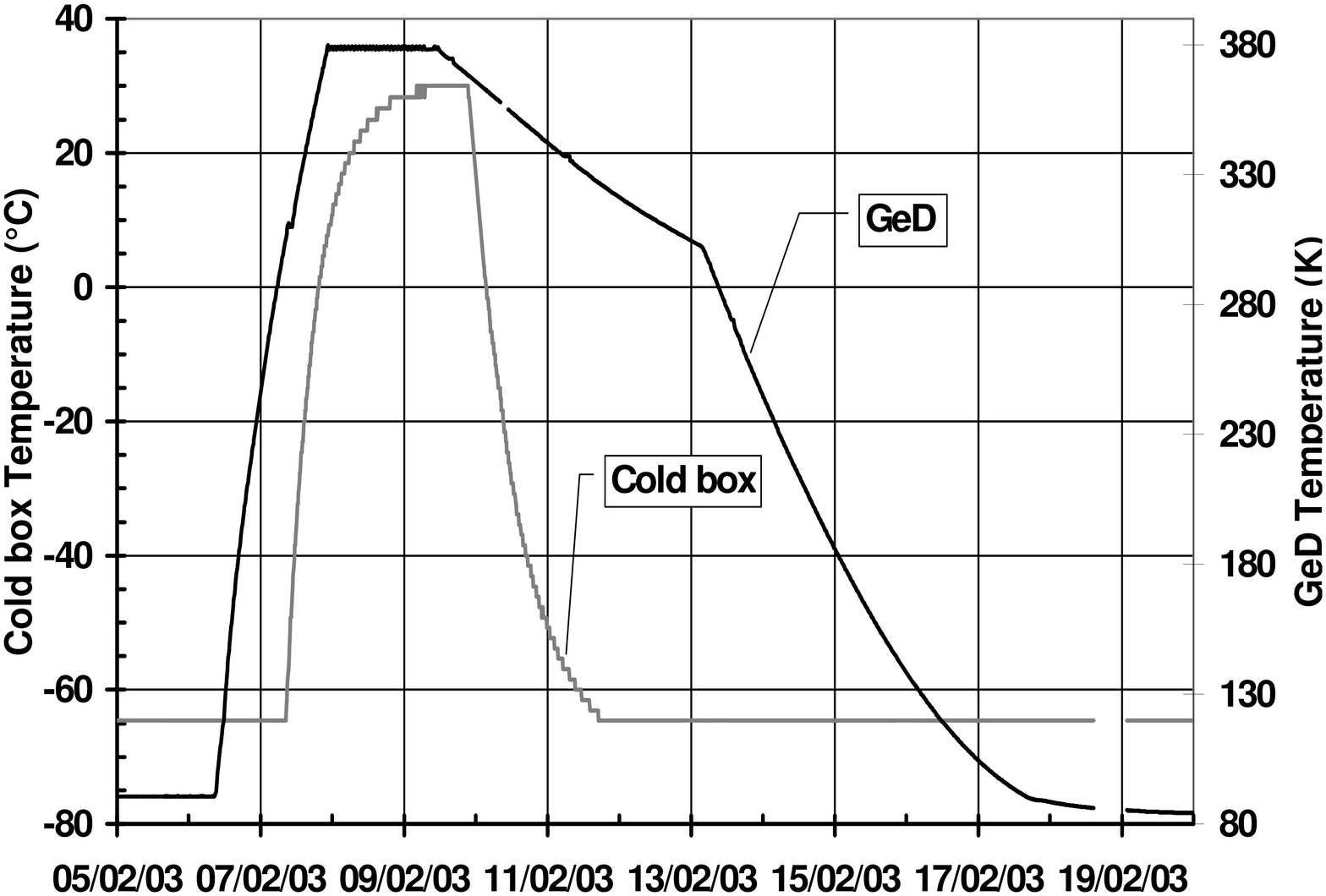}
\caption{Temperature profiles during annealing.}
\label {therm2}
\end{figure}

The timing of the first SPI annealing was decided based on the following arguments:
\begin {itemize}
\item We knew (GeD 15) that some contamination exists within the GeDs.
\item The GeDs exhibit a significant degradation.
\item  We wanted to have nominal GeDs for the Crab observation that took place immediately after the annealing.
\end {itemize}
The annealing process is summarised in figure \ref {therm2}. The GeDs were maintained at 105.7$^\circ$C during 37 hours while the cryostat envelope was maintained at 28$^\circ$C during 28 hours. All operations were smoothly executed and normal operations were resumed after 12 days.
Table 1 shows the mean energy resolution for 4 background lines of the whole camera (excluding GeD 15 because it improved also due to outgassing). The recovery was nearly perfect. The annealing capability of SPI was a constraint on the design of the camera but was necessary to ensure good performance for a 5-year lifetime mission.
The second annealing was performed from 2003 July 18 to 30.

\begin{table}
\caption[]{Mean energy resolution for different epochs.}
\label{tbl-1}
\begin{tabular}{ccccc}
\hline
\noalign{\smallskip}
Date & 198keV& 882keV & 1778keV& 2754keV\\
\noalign{\smallskip}
\hline
\noalign{\smallskip}
After launch               & 1.82 & 2.33 & 2.92 & 3.87 \\
02/12/31                   & 1.79 & 2.34 & 3.18 & 4.4 \\
Before annealing           & 1.85 & 2.47 & 3.34 & 4.57 \\
After annealing            & 1.83 & 2.32 & 2.97 & 3.91 \\

\hline
\end{tabular}
\end{table}

\subsection {Energy calibration in the 20-2000 keV energy range}

\begin{figure} [here]
\vspace*{-0.7cm}
 \centering
  \includegraphics[width=6.5cm,angle=270]{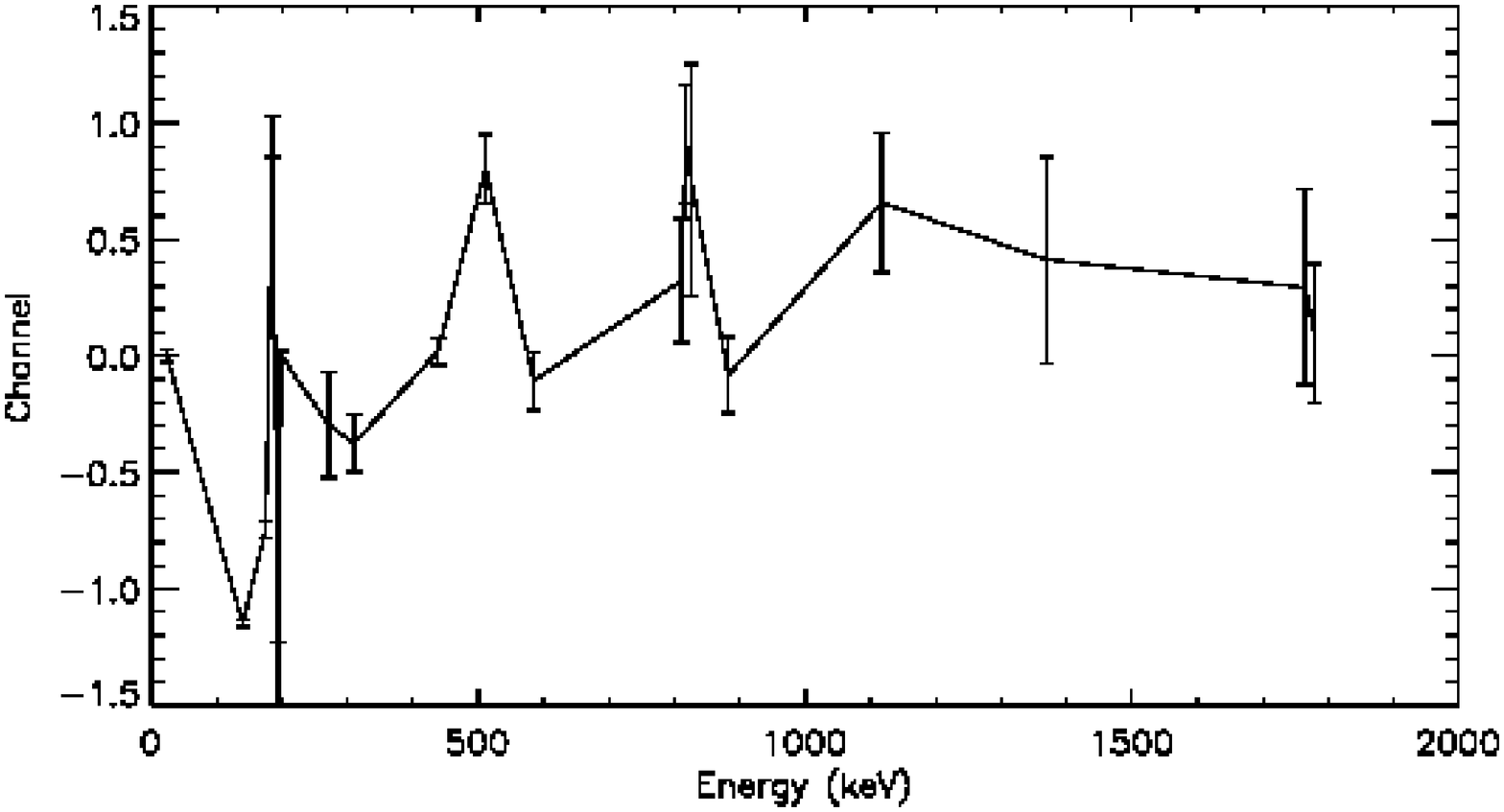}
\vspace*{-1.5cm}
\caption{Residuals between measured peak positions and predicted energy.}
\label {linearity}
\end{figure}

For this study the following reference lines have been chosen: 
23.43 keV from $^{71m}$Ge (this line is emitted together with a line at 174.9 keV thus, in order to avoid the blend with the 24.8 keV line from $^{58m}$Co, the 23.43 events are taken from the multiple events);
198.34 keV from $^{71m}$Ge; 438.6 keV from $^{69m}$Ge; 882.35 keV from $^{69}$ Ge; 1778.9 keV from $^{28}$Al.
Using these line energies and the corresponding line positions in the spectra a third order polynomial function was fitted. Then, 20 line positions were compared with the predicted values. The residuals for GeD 0 are plotted in figure \ref {linearity}.

 The residuals show that using this scheme the low energy range can be calibrated with an accuracy of $\pm$1 channel at worst, one channel corresponds to 0.135 keV. If a simple linear function is used, the maximum error found is 3 channels i.e. 0.4 keV.
The results are similar for all 19 electronics chains. 

\subsection {Gains drifts}
From ground tests it has been shown that the gain drift due to the pulse shape amplifier and pulse height analyser temperature is negligible, furthermore the preamplifier temperature is extremely stable (see section 2.4). Thus the main contribution to the gain drifts should originate in temperature changes of the cold plate. The changes of gain were measured during the cooling down phase. At 1117 keV the drift is 0.187 keV/K between 90 and 95 K. The cold plate temperature, controlled from ground, has always been stable to $\pm$1 K. Direct gain correction using the GeD temperature is under study. If feasible the gain calibration will be highly simplified.

%%%%%%%%%%%%%%%%%%%%%%%%%%%%%%%%%%%%%%%%%%%%%%%%%%%%%%%%%%%%%%%%%%%%%%%%%%%%%%%%
% PSD
%%%%%%%%%%%%%%%%%%%%%%%%%%%%%%%%%%%%%%%%%%%%%%%%%%%%%%%%%%%%%%%%%%%%%%%%%%%%%%%%
\section{PSD}
\label{sec:psd}

From a functional point of view, the PSD sub-assembly is basically working 
as expected.

The overall trigger rate of the sub-assembly amounts to $300-500$ Hz, well below the maximum processable trigger rate of about 1 kHz.
Thus all registered events can be analysed by the PSD sub-assembly within the available processing time.
About 55-70\% of the PSD triggers are noise triggers, i.e. triggers that are not related to a real Germanium detector event.
Although these events are analysed as regular events, they are dropped later due to the absence of corresponding information from the AFEE sub-assembly.
From the pulse shapes that have been registered for the noise triggers it appears that they are probably related to baseline current instabilities. 
These instabilities are likely caused by energetic particle interactions within the Germanium detectors, i.e. by particles that deposit energies well above the upper level discriminator threshold of the PSD ($\sim2.7$ MeV).
This hypothesis is substantiated by the fact that the activation of the pre-amplifier clamping circuit reduces the number of noise triggers (from 70\% to 55\%).

The maximum PSD efficiency, defined as the fraction of real Germanium detector events that triggered the PSD with respect to the events that triggered the AFEE (energy chain), is reached in an energy band from about 400 keV to 1.5 MeV and amounts to 85\%. 
This means that the PSD sub-assembly misses about 15\% of all events in the PSD energy band. 
It appears that this loss is due to baseline instabilities that occur due to energetic particle interactions within the Germanium detectors. 
The front-end setting (notably the FET threshold) has a crucial impact on the PSD efficiency, and has been optimised in order to maximise the PSD efficiency.

%The spurious events that are seen by the AFEE sub-assembly around 1.5 MeV are %not detected by the PSD sub-assembly, hence using the PSD trigger information %may be used to reduce this artefact.

The primary goal of the PSD sub-assembly consists in rejection of single-site 
Germanium detector events while conserving at the same time multiple-site events. 
Single-site events are primarily created by local radioactivity within the detector volume, hence are attributed to instrumental background. 
Multiple-site events primarily arise from photon interactions within the Germanium detectors, hence they are attributed to celestial photons. 
The effective sensitivity improvement that may be achieved by this discrimination technique strongly depends on the precise nature of the instrumental background, in particular the fraction of localised or single-site events (\cite {gehrels1985}).
From pre-launch simulations, this fraction was estimated to about 95\%, 
resulting in a theoretical sensitivity improvement of about a factor of 2.

From the analysed in-flight data, the fraction of localised or single-site 
events had to be revised downwards to about 80\%. 
With such a low fraction, the achievable PSD sensitivity improvement is 
considerably reduced. 
The analysis of instrumental background lines has suggested sensitivity improvements in the $10-30\%$ area. Application of the PSD to Crab nebula data has not shown any significant sensitivity improvement.
Thus it appears that the low fraction of single-site events in the instrumental background (95\% expected, 80\% observed) inhibits a significant sensitivity improvement by the PSD sub-assembly. 
In other words, the characteristics of the instrumental background and the celestial signal are too similar and do not allow for an efficient pulse shape discrimination.

\section{Tuning and testing of ACS performance}

During the commissioning phase the ACS has been tuned and calibrated and several investigations were performed in order to improve the performance and the scientific understanding of the ACS. Many results of these investigations were used to improve the background simulations (\cite{pjean2003}).

\subsection{ACS calibration}

\begin{figure}
   \centering
   \includegraphics[width=0.50\textwidth,angle=0,clip]{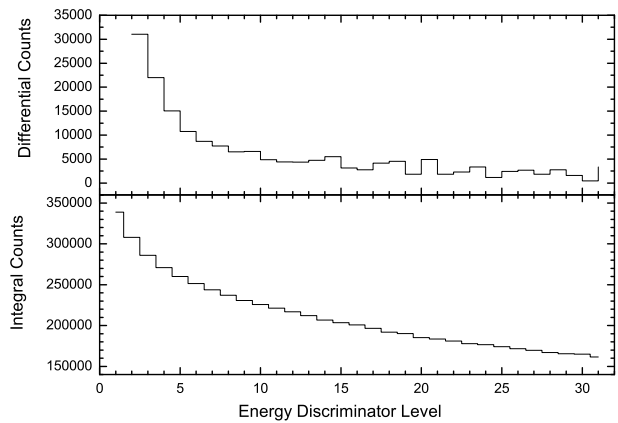}   
\caption {Integral (bottom) and differential (top) spectra for FEE0.
Extracted from ACS calibration}
         \label{fig:GRB-loc} 
   \end{figure}

The ACS calibration was performed by monitoring the dependence of the individual FEE count rate on the energy-discriminator level. The whole commandable range is covered in steps of two threshold levels starting from level 0 (78 mV) up to level 63 (4.9 V). At each setting the FEE veto rate was measured (the counter was set to 2 s measurement time). The ACS calibration is an automatic procedure of the ACS on-board software. 46 repetitions of this procedure were necessary in order to gain enough statistic. The measurement is merged into an integral spectrum for each FEE. Figure \ref {fig:GRB-loc} shows the result for the ACS upper ring. By calculating the difference in count rate between adjacent steps one gets a differential spectrum for each FEE. 

These spectra didn't show any significant feature like the 511 keV line, but the measured dependencies can be used for the improvement of simulations. The dependencies were also used for the tuning of the energy-discriminator levels described below. Monitoring the ACS overall counter during the ACS calibration gives information on the "dynamic range" of the ACS. Changing the energy discriminator of all FEEs from the nominal 100 keV setting to its maximum value (roughly 300 keV), the ACS overall count rate is approximately halved. 

%------------------------------------------------------------------
%------------------------------------------------------------------

\subsection{Tuning of the ACS energy thresholds}

%------------------------------------------------------------------
  \begin{figure}
   \centering
  \includegraphics[width=0.33\textwidth,bb= 50 150 550 670,angle=270]{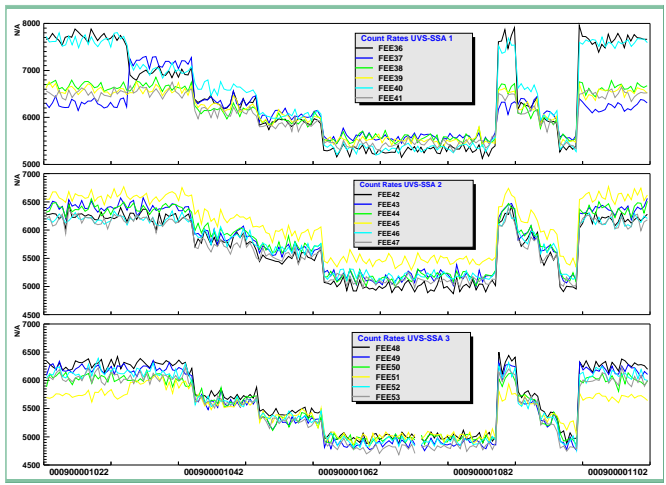} 
\vspace{0.5cm}
\caption {Tuning of the ACS energy threshold. From left to right: 100\,keV untuned, 100\,keV tuned, 150\,keV tuned,
200\,keV tuned, 300\,keV tuned. And for comparison again the untuned setting for: 100\,keV, 150\,keV, 200\,keV, 300\,keV and 100\,keV}
         \label{fig:E-thres} 
   \end{figure}

%------------------------------------------------------------------

One important change compared with the on-ground conditions is the lower than expected mean temperature of the ACS. This affects the gain of all detection chains, and thus of the energy threshold of the ACS. Compared to the conditions during the on-ground calibration of about 25$^\circ$C the new temperature mean range spans between -1.1$^\circ$C and -8$^\circ$C. An adaptation of the on-ground settings was therefore necessary.

The scientific performance of the ACS is determined by the setting of the energy-discriminator level and the accuracy of the timing. The low-energy threshold was tuned in such a way that the FEEs which were connected to crystals with a similar size and shape measure approximately the same counting rate. This has been performed for 75, 100, 150, 200 and 300 keV thresholds. This goal was achieved and the counting rates had a spread of less than 5\% 
(Figure\,\ref{fig:E-thres}). 
The energy threshold was then set at a value of $\sim 75$\,keV. Because of the redundancy concept of the ACS (the rates of two PMTs viewing two different crystals with different light yields are summed in one FEE) and the different amplifications of the PMTs the energy threshold is poorly defined (i. e. it is not a sharp step function, but a smooth increasing function extending from $\sim 50$\,keV to $\sim 150$\,keV). No upper energy limit for high-energy deposits exists, i. e. even the events with very high energy deposits are counted. However in order to recognise the number of high-energy events a so-called over-range threshold exists which counts only those events which exceed this threshold. The value of this energy threshold is only very coarsely known. It lies somewhere between $\sim 50$\,MeV and 150\,MeV. The actual over-range counting rate measured with the DFEE is $\sim 6000$\,counts/s.

Considering the response of the whole SPI, a nominal ("best") ACS setting of the energy-discriminator levels was chosen. A detailed analysis of the germanium-camera background recommended a 75 keV threshold for the whole ACS. With this setting the ACS overall count rate is $\sim 66000$\,cts/s and the background of the GeDs is reduced by a factor 20.

%------------------------------------------------------------------
%  \begin{figure*}[t]
%   \centering
%   \includegraphics[width=0.4\textwidth,bb= 250 44 571 754,angle=270]{ACS-%100d_030606.ps}   
%\caption {Long-term stability of the ACS}
%         \label{fig:ACS-long} 
%   \end{figure*}

%------------------------------------------------------------------

\subsection{Timing of the ACS}

The timing of the ACS overall counter has been checked and aligned with the universal time (UT). The overall counting rate is measured with 50 ms time bins. The start time of each bin is known with respect to the UT with an accuracy of $\pm 10$\,ms. This knowledge is important, because it is required in the calculation of the arrival direction of a gamma-ray burst (GRB). The onset of the GRB lightcurve is an input to the interplanetary network (IPN) and the ACS data are used to localise GRBs down to an accuracy of a few arcminutes (\cite{kienlin2003}).

\subsection{Leakage of the ACS}

One important criterion to judge the quality of the ACS is the
electronics leakage rate. This rate is related to the dead time of the
ACS: after each event the ACS registers, the electronics of the ACS is
for a certain time not able to register another event. If this time
exceeds the length of the veto signal, no new veto will be generated
until the recovery of the electronics. This effect is leading to a
leakage, which means a non-vetoing of GeD events, especially for
overrange events where the electronical dead time is long. A test of the
shield leakage was performed by halving the gains of all ACS
photomultipliers via a reduction of the corresponding HV and an adapted
reduction of the energy-discriminator levels. The idea was to reduce the
electronic dead time of the front-end electronics by reducing the amount
of charge produced by each overrange event. The event rate in the GeD
detectors remained unchanged, showing that the contribution of the
electronic leakage of the ACS to the Ge-detector background is small or
even negligible.

In a test with the PSAC (switched on and off) it was found that the PSAC makes only a minor contribution to the reduction of the background. For the 511 keV a reduction of about 5\% was observed.

\subsection{Long term stability of the ACS}

During the commissioning phase the ACS has been tuned and calibrated and has worked perfectly since then, measuring an overall background counting rate of $\sim 66000$\,counts/s. 
In order to investigate if the threshold levels remain stable the ACS overall rate has been monitored for a time period of 100 days. If we except solar flares and radiation belts entries, the counting rate is stable within $\sim$10\%.
%Fig.\,\ref{fig:ACS-long}.

% ============================================================================
\section {Internal timing performance}
% ============================================================================
        The SPI-system level tuning-phase took place during the PV-phase. 
        The interactions between the Ge-camera, ACS, PSD and DFEE (\cite{sch2002}) 
were adjusted as during on-ground tests (\cite{sch2003}).

        The signal timing alignment at the level of the DFEE association machine 
(\cite{laf1998}, \cite{mur2002}), where the SPI event-building takes place has been checked in flight. 
        The delays of the DFEE input stage were adjusted such that time-tags issued by two Ge-detectors, corresponding to a same physical event (a photon,
Compton scattered in one detector, and absorbed in an other), arrive 
simultaneously at the association machine input. 
        Such an event is classified as ME (Multiple Event). 
        Furthermore the association time window for ME building is optimized to cover fluctuations in time-tag arrival due to electronics jitter, but kept small enough not to group many accidentals in ME.
        
\begin{figure}
\centering
  \includegraphics[width=8.8cm,angle=0]{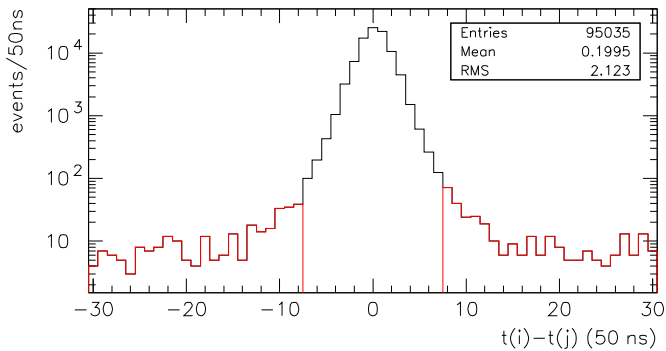}
\caption {
        Distribution of the time difference between hits on any neighboring detector pair $(i,j)$ using the first 2 hits of multiple events.
        The mean misalignment between detector pairs is 10 ns. 
        An association time window of 350 ns has a ME association inefficiency of only 1\% for SPI in-flight events (mostly background).
}
\label{label_dtme}
\end {figure}

        For the purpose of this in-flight check, the DFEE association time window is set wide open and the in-flight ME are analyzed. 
        Using the first 2 hits of a ME, the distribution of the arrival time difference $t(i)-t(j)$ for each pair $(i,j)$ of neighboring detectors is evaluated.
        The DFEE input delays are adjusted such that all detector pairs are aligned within the DFEE timing accuracy of 50 ns (Figure \ref{label_dtme}).
        The association time window is set to 350 ns in order to reduce the accidental ME association.
        Only 1\% of the previous ME are now classified as SE.

        An other measurement using PE (single Ge-detector hit with PSD trigger) permits to check successfully that the alignment of the PSD time-tag with respect to the Ge time-tag is better than 50 ns and the association time window of 350 ns is large enough to group all PSD and Ge time-tags inside PE.

\begin{figure}
\centering
\includegraphics[width=8.8cm,angle=0]{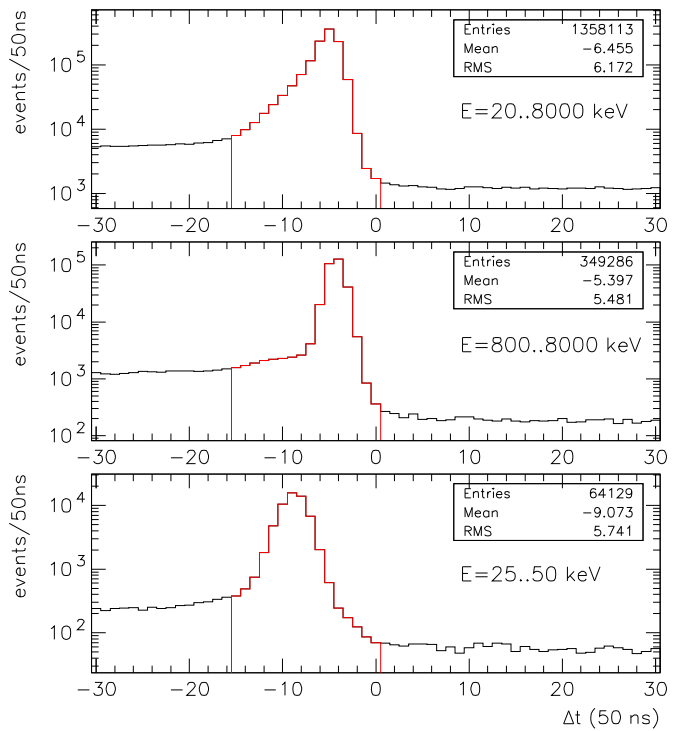}
\caption {
        Distribution of the difference $\Delta t=t(ACS)-t(Ge)$ between the arrival time of an ACS time-tag and a time-tag from any of the 19 Ge detectors (top graph).
        The signals are well aligned in time: all non-accidental hits show $\Delta t<0$, meaning that the ACS trigger arrives before the Ge detector time-tag.
        The ACS gate width is well set to 750 ns: all non-accidental hits have $|\Delta t|<$750 ns, even considering the time-tags at the extreme of the energy range: middle and bottom graphs.
This width is kept as low as possible for a low ACS dead-time (8\% with 
this setting).
}
\label{label_dtacs}
\end {figure}

        In order to perform the alignment of the ACS gate with respect to the Ge time-tags, we used events triggering both the ACS and a Ge-detector and we measured their time difference $\Delta t$ (Figure \ref{label_dtacs}). 
        We conclude that all ACS triggers arrive before the Ge time-tags and that for an efficient veto function and a minimal veto dead-time, the length of the ACS veto gate must be set to 750 ns. 
        For this purpose, the DFEE adds to an incoming ACS veto gate (200 ns wide) a gate-extension (set to 550 ns) before its usage in the veto function.

% ============================================================================
\section {SPI counting rates} 
% ============================================================================
In flight, a typical SPI Ge detects about 950 non saturating hits/s.
Each of the corresponding time-tags (dead-time of 27 $\mu$s for ADC integration) is sent to the DFEE together with its ADC energy value.
        Additionally about 190 saturating time-tags (events of energy above 8 MeV) are produced (100 $\mu$s dead-time each, without energy value).

\begin{figure}
\centering
  \includegraphics[width=6cm,angle=0]{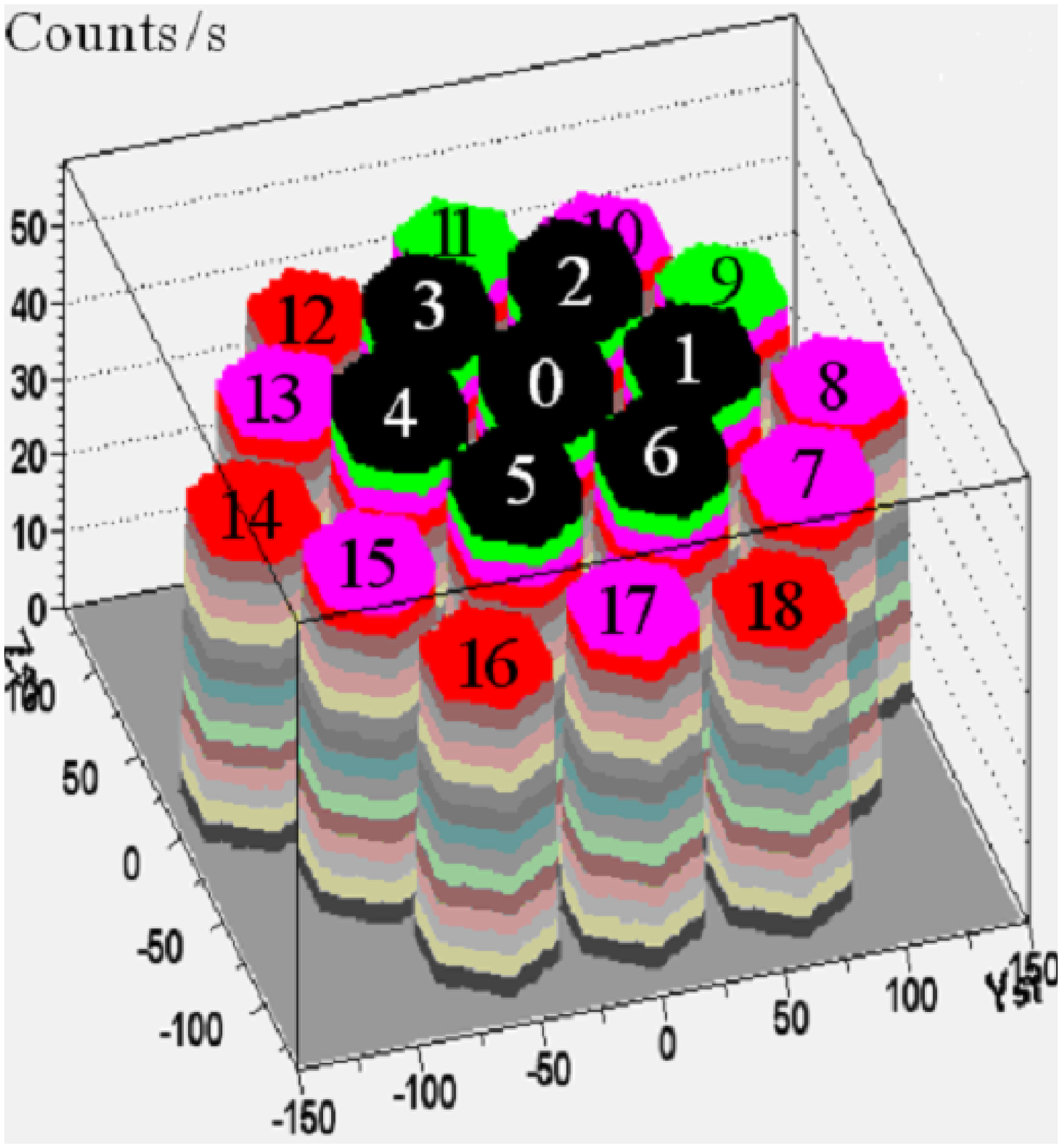}
\vspace*{-1cm}
\caption {
        Distribution of the non saturating, non vetoed Ge counts on the SPI camera (vertical axis in counts/s, Ysi and Zsi axis=positions of the detectors in mm, detector numbers marked on the plot).
        The outer detectors show less counts, since Compton scattered photons may escape into the BGO shield and activate a veto signal, which suppresses the event (self-veto effect).
}
\label{label_cnts}
\end {figure}

\begin{figure}
\centering
\includegraphics[width=8.7cm,angle=0] {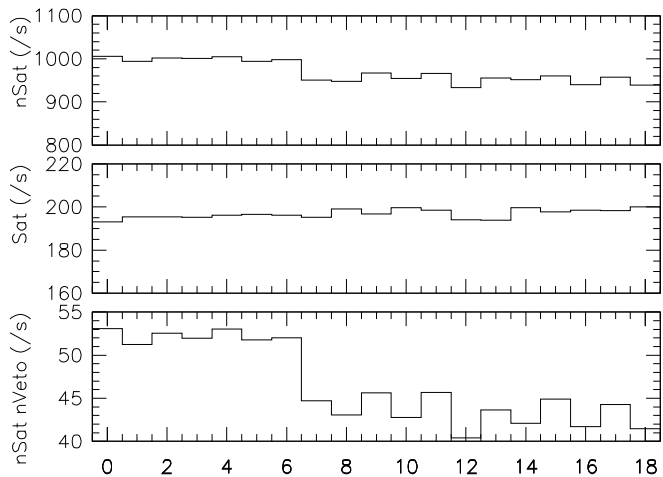}
\caption {Count rates observed for GeDs 0 to 18 for: all counts below 8 MeV (top graph); saturating events (energies greater than 8 MeV, middle graph); and non vetoed counts below 8 MeV (bottom graph).}

\label {figcounts}
\end {figure}

        Particles exceeding a threshold of about 75 keV, interacting in the BGO scintillator of the ACS produce veto signals sent to the DFEE, at a rate of about 66000/s.
        The DFEE applies the veto function and reduces the 950 non saturating Ge time-tags/s to about 50 non vetoed and non saturating Ge time-tags/s (figures \ref{label_cnts} and \ref{figcounts}).
        Those are used for event building and classification into SE, ME and PE.        Each time-frame (of 125 ms), 53 SE, 20 PE and 12 ME are produced. 
        This event share is obtained for a PSD energy range set to 300-2000 keV. 
        With a higher PSD energy threshold, the balance is shifted towards more SE and fewer PE, which spares some telemetry.

        The 66000 veto counts/s are composed of (potentially overlapping) non saturating veto signals (65300/s, duration 750 ns each) and saturating veto signals (5800/s, duration 5.55 $\mu$s each, 2.6 $\mu$s for the ACS gate and additional 2.95 $\mu$s for the internal DFEE 
extension).
        Those veto signals, used by the DFEE for event rejection, introduce a veto 
dead-time of 8\%.
        Combined together with the dead-time for non-saturating and saturating Ge time-tags, each Ge detector has a global dead-time of about 12\%. 
        This dead-time is measured by the DFEE, counting the fraction of time when the veto signal OR the non-saturating OR the saturating time-tag signal of a given Ge detector is active.

        With these in-flight counting rates the global SPI telemetry need is 93 telemetry packets (of 440 bytes) per 8 s, most of which (80 packets) is used for transmission of the scientific data (SE, ME, PE). 
        However before launch, only 36 packets were allocated to SPI. 

        During in-flight tuning, unsuccessful attempts have been made to get an even better rejection factor for the ACS (the 950 time-tags/s reduced to even less than 50). 
        By increasing the veto signal duration (non saturating, as well as saturating veto signals separately), we hoped to reject many background events from BGO activation.
        However, we observed only a decrease in the non-vetoed event counting rate proportional to the increased dead-time.
        Spectral analysis showed that many background events come from activation of the Ge-detectors themselves and cannot be rejected by the ACS.

% ============================================================================
\section {SPI telemetry allocation}
\begin{figure}
\centering
  \includegraphics[width=8.8cm,angle=0]{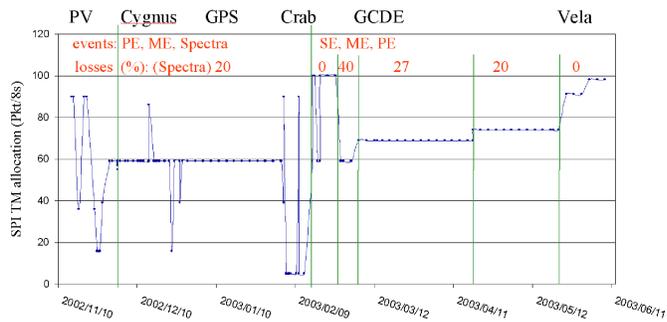}
\caption {
        Evolution of the SPI telemetry allocation with time and 
corresponding data losses.
}
\label{label_tm}
\end {figure}

% ============================================================================
        For a long time SPI was operated with less than the 93 packets/8s needed, resulting in a significant amount of data loss (or observing time, figure 
\ref{label_tm}). 
        After telemetry margin discovery at INTEGRAL level, and successful negotiations within the INTEGRAL team, the SPI telemetry allocation is now 98 packets/8s, giving SPI its full efficiency.

        In the future, closer to the minimum of solar activity, the counting rates are expected to increase and the SPI telemetry allocation will again be insufficient. 
        The SPI team is preparing a reduction of the telemetry need by changing the 
on-board software. 
        In the on-board ME table all events with multiplicity$>3$ (scientifically not of high importance) are rejected, the $\Delta t$ information is suppressed, and the timing accuracy is degraded (from 102.4 $\mu s$ to 409.6 $\mu s$ for double ME, and 819.2 $\mu s$ for triple ME). 
        In this way, double ME will use 6 bytes and triple ME 8 bytes of telemetry, and the SPI telemetry need is reduced by about 11 packets/8s. 
        Further telemetry reduction can be achieved by instrumental background line-filtering of the SE table and is presently under investigation.

\section {Imaging performance}

The SPI imaging performance is highly dependent upon the analysis method, the complexity of the field of view (number, relative intensity, flux stability of sources) and the background handling. We have seen (section 7) that the background is not uniformly distributed on the camera. This shape, that can be determined from empty field observations for each energy bin, has been found relatively stable. These "uniformity maps" can be scaled by means of one free parameter for a given time interval. The time interval depends on the background stability and can range from a few hours to a few days. The results shown here are obtained using such background modelling, the images are build through the SPIROS software (\cite{ski2003}). The first SPI images have been obtained during the observation of the Cygnus region; these results are discussed in \cite{bou2003}.
\subsection {Crab Nebula observation}
\begin{figure} [here]
 \centering
\vspace*{-0.7cm}
  \includegraphics[width=8.0cm,angle=0]{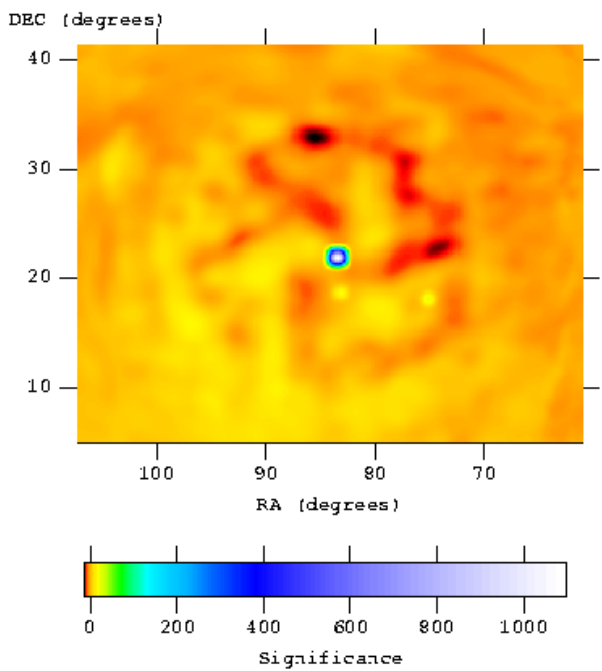}
\caption{Image of the Crab Nebula in the 20-50 keV energy range, the scale is logarithmic.}
\label {crabfig}
\end{figure}

Figure \ref {crabfig} shows the reconstructed image in the 20-50 keV energy range using 7.54 days of data. The source is detected at 1097 $\sigma$ while the strongest ghosts have an intensity of 1.5 \% of the maximum. The defects shown here are dominated by the precision of our imaging response. 

The uniformity of the response has been measured by dividing the data into subgroups with approximately the same pointing offset angle from the Crab. Up to 12 degrees off-axis the reconstructed source intensity is stable within a few \%. Above 12 degrees stronger deviations appear (20-30\%), that is not understood at this time. Any source extraction at the edge of the field of view requires carefull analysis.
\subsection {Galactic plane}

\begin{figure*}
 \centering
  \includegraphics[width=17.0cm,angle=0] {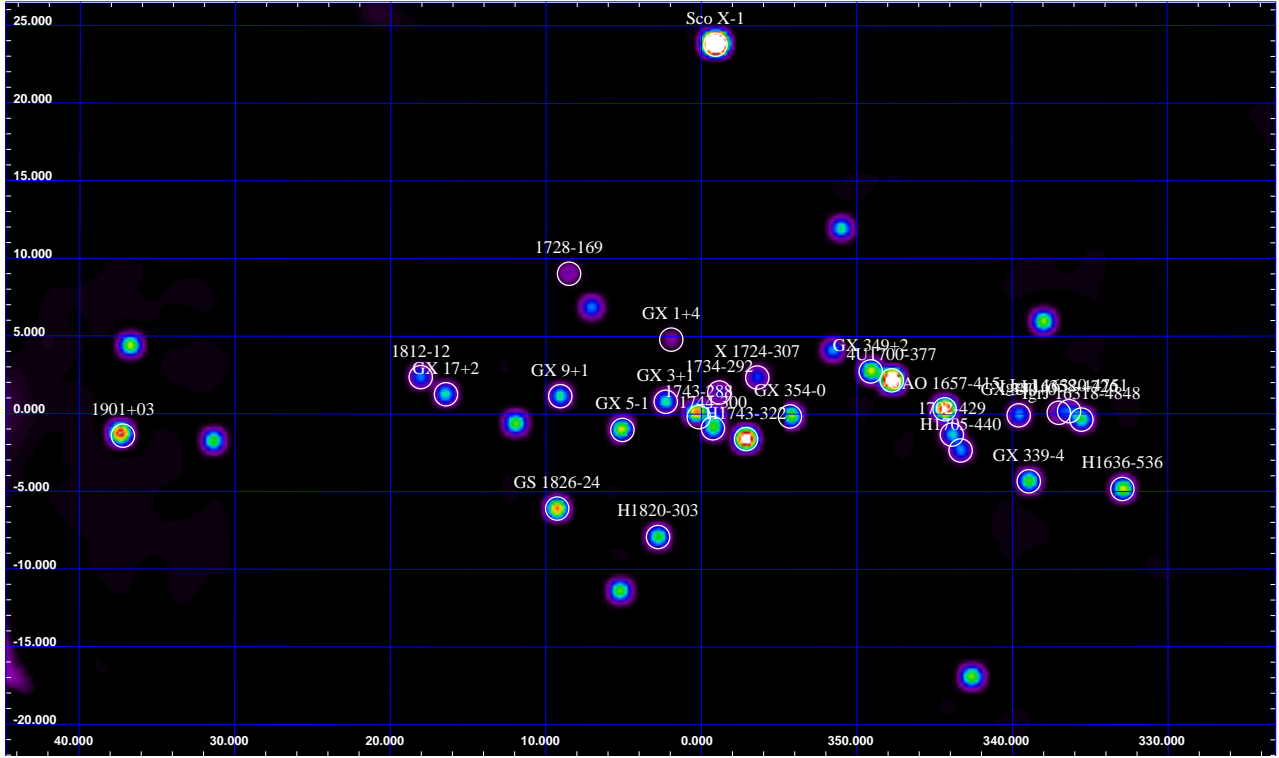}
\caption{Image of the Galactic plane obtained during the Galactic Center Deep Exposure 20-50 keV.}
\label {galfig}
\end{figure*}

The Galactic plane is an example of a highly complex region that has been extensively observed by INTEGRAL. The data used here spans from revolution 47 to revolution 65. 703 pointings of ${\sim}$ 30 min have been used which correspond to 15.4 days livetime. Figure \ref {galfig} shows the result obtained in the 20-50 keV energy range. 

\section {Spectral reconstruction}
\begin{figure} [here]
 \centering
\vspace*{-0.7cm}
  \includegraphics[width=9.0cm,angle=0]{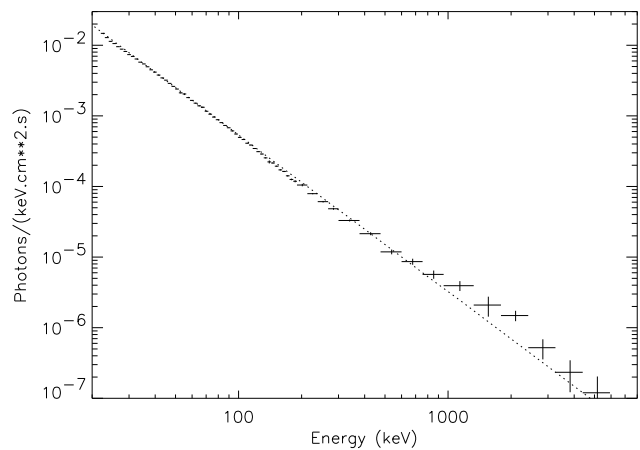}
\caption{Spectrum of the Crab nebula.}
\label {crab-spectrum}
\end {figure}

The Crab region has been observed during 6.57 days for calibration purposes. Figure \ref {crab-spectrum} presents a preliminary spectrum of the Crab nebula reconstructed using the latest version of our response (July 2003 version). A power law fit gives the following result: Flux at 100 keV: 0.536 10$^{-3}$ ph/cm$^{2}$/s/keV, photon index: 2.23. A fit with a broken power law gives a power law index of 2.02 below 50 keV and 2.29 above 50 keV. The general shape of the spectrum gives a qualitative idea of our response quality; see also \cite{sturner2003} and \cite{att2003}.

\section {Sensitivity}
\begin{figure} [here]
 \centering
  \includegraphics[width=8.5cm,angle=0]{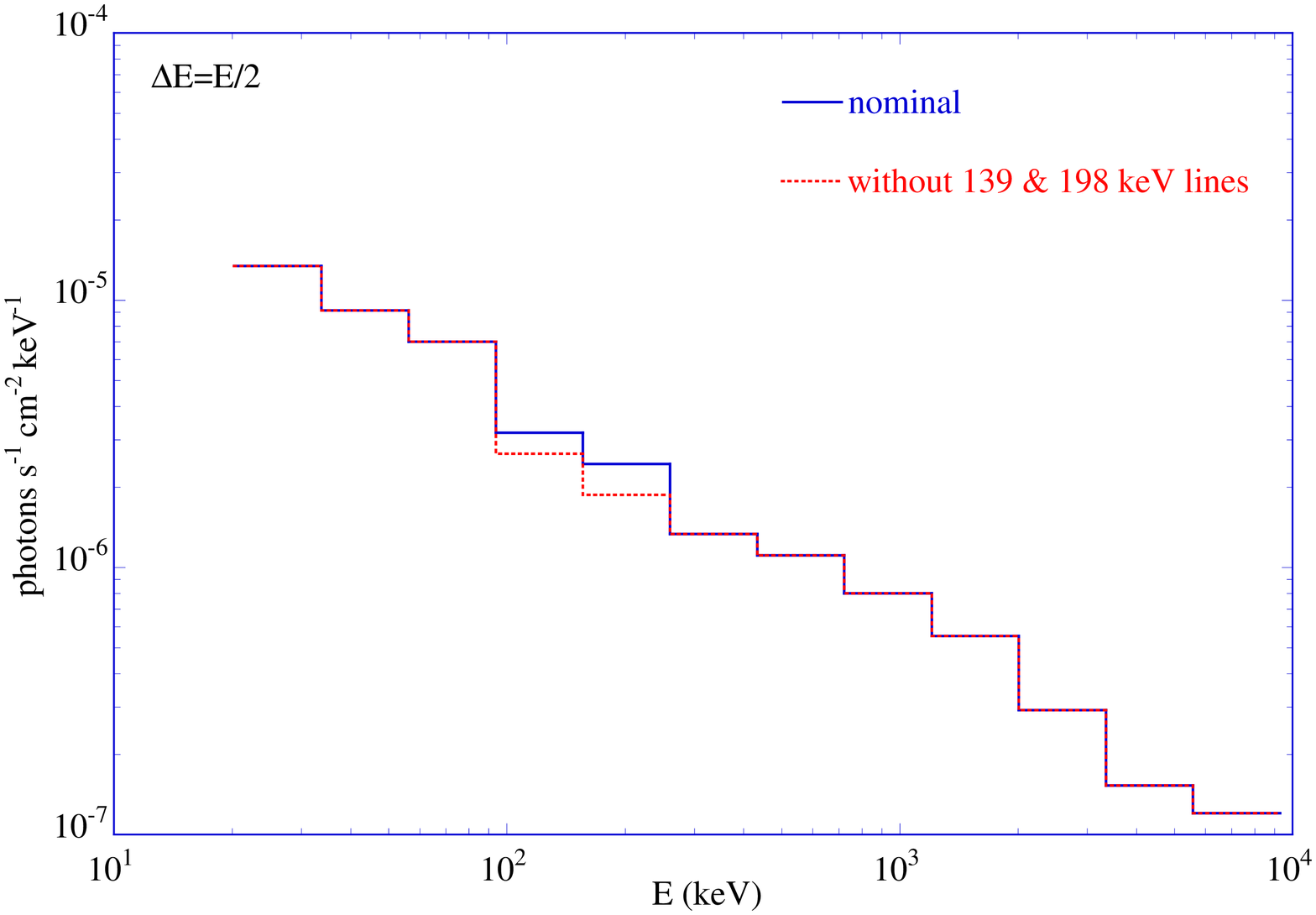}
\caption{Continuum sensitivity for on-axis point source and for 10$^6$s observation time.}
\label {conti-sens}
\end {figure}

\begin{figure} [here]
 \centering
  \includegraphics[width=8.5cm,angle=0]{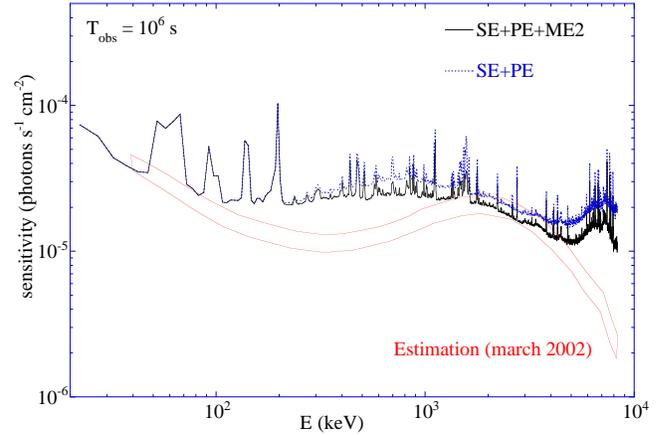}
\caption{3 $\sigma$ narrow line sensitivity for 10$^6$s observation time.}
\label {line-sens}

\end{figure}

The 3 $\sigma$ sensitivity limit of the instrument has been computed for 10$^{6}$s of observation time for the continuum (figure \ref{conti-sens}) and for narrow lines (figure \ref{line-sens}). The quoted sensitivities only include the statistical errors: the impact of imaging algorithm and background susbtraction methods is not included. The sensitivity limit is 5 10$^{-5}$ph$\cdot$cm$^{-2}\cdot$s$^{-1}$ at 511 keV and 2.5 10$^{-5}$ph$\cdot$cm$^{-2}\cdot$s$^{-1}$ at 1809 keV.

%%%%%%%%%%%%%%%%%%%%%%%%%%%%%%%%%%%%%%%%%%%%%%%%%%%%%%%%%%%%%%%%%%%%%%%%%%%%%%%%
% Acknowledgements
%%%%%%%%%%%%%%%%%%%%%%%%%%%%%%%%%%%%%%%%%%%%%%%%%%%%%%%%%%%%%%%%%%%%%%%%%%%%%%%%
\begin{acknowledgements}
The SPI project has been completed under the responsibility and
leadership of CNES. We are grateful to ASI, CEA, CNES, DLR, ESA, INTA,
NASA, and OSTC for support.
\end{acknowledgements}

\end{document}